\documentclass[pra,superscriptaddress,onecolumn,longbibliography,nofootinbib]{revtex4-2}

\usepackage[T1]{fontenc}
\usepackage[utf8]{inputenc}
\usepackage{physics}
\usepackage{amsmath}
\usepackage{amssymb}
\usepackage{mathtools}
\usepackage{tensor}
\usepackage{hyperref}
\hypersetup{
    bookmarksopen=false,
    bookmarksnumbered=true,
    pdfnewwindow=true,
    unicode=false,
    colorlinks=true,
    citecolor=blue,
    linkcolor=black,
    urlcolor=blue,
    filecolor=blue
    }
\newcommand{\usal}{Departamento de F\'isica Fundamental, Universidad de Salamanca, E-37008 Salamanca, Spain}
\newcommand{\iffym}{Instituto Universitario de F\'isica Fundamental y Matem\'aticas (IUFFyM), Universidad de Salamanca, E-37008 Salamanca, Spain}

\begin{document}

\title{Dynamical Behaviour of Density Correlations Across the Chaotic Phase for Interacting Bosons}

\author{Óscar Dueñas}
\email[]{osdusa@usal.es}
\affiliation{\usal}
\affiliation{\iffym}
\author{Alberto Rodríguez}
\affiliation{\usal}
\affiliation{\iffym}


\begin{abstract}
    We investigate the propagation of two-point density correlations in the one-dimensional Bose-Hubbard Hamiltonian in the thermodynamic limit 
    in terms of the correlation transport distance (CTD), an experimentally measurable magnitude that characterizes the spatial spreading of correlations in time. 
    We confirm that the integrable limits of the model exhibit CTD ballistic growth, while the onset of the chaotic phase leads to the emergence of a pronounced sub-ballistic regime, in agreement with previous results for finite systems. 
    By a meticulous analysis of the spatio-temporal correlation profiles, we show that the correlation front nonetheless propagates ballistically 
    for all interaction strengths, and that the chaos-induced slowdown of the CTD 
    originates from the emergence of long-time distance-dependent correlation tails, together with an enhanced decay of the correlation front amplitude. Our results thus provide a detailed characterization of correlation transport that goes beyond a simple light-cone picture.
\end{abstract}

\maketitle
\section{Introduction}
In quantum many-body systems, the interplay between many-particle interference and interactions may give rise to a rich structure of phases, potentially including a chaotic (ergodic) regime. 
While signatures of chaos are commonly identified through spectral statistics and eigenstate structure \cite{Haake2018,Izrailev1990}, via the benchmark of these against random matrix theory 
\cite{Brody1981,Guhr1998}, the existence of quantum chaos also leaves a recognizable imprint in the dynamical response of the closed system \cite{Deutsch1991,Rigol2008,Borgonovi2016} which is typically more accessible to experiments, particularly with platforms of ultracold bosons in low dimensional optical lattices 
\cite{Cheneau2012,Gring2012,Trotzky2012,Ronzheimer2013,Langen2013,Meinert2014,Langen2015,Kaufman2016,Bordia2016,Choi2016,Lukin2019,Rubio-Abadal2019,Rispoli2019,Leonard2023,Wienand2023}.

Under the premise of a chaotic regime in the excitation spectrum, local observables (i.e., those involving a reduced number of degrees of freedom) approach an equilibrium expectation value in time characterized by temporal fluctuations that decrease exponentially with the number of degrees of freedom of the system \cite{Srednicki1996,Srednicki1999}. How this dynamical ergodicity emerges is nonetheless crucially affected by the nature of the initial state considered \cite{Pausch2025}. Additionally, quantum chaos may also affect the spreading of information across the system, assessed by monitoring the time evolution of many-particle observables such as experimentally measurable density-density correlations \cite{Cheneau2012,Rispoli2019,Wienand2023}. In fact, recent experimental investigations suggest a hydrodynamic approach to the dynamics of many-particle correlations in the chaotic regime \cite{Wienand2023}.

In a former study \cite{Duenas2025}, upon introducing a convenient two-particle correlation transport distance, we demonstrated that the chaotic phase in a system of interacting bosons in a finite one dimensional lattice, modelled by the Bose-Hubbard Hamiltonian, induces a sub-ballistic, arguably diffusive, regime in the propagation of two-point density correlations. This result seems at odds with previous theoretical and experimental observations for one dimensional bosons, where the spreading of density correlations was reported to follow a ballistic light-cone for any value of the interparticle interaction strength \cite{Cheneau2012,Lauchli2008,Barmettler2012,Despres2019}.

The purpose of this work is to resolve this apparent discrepancy by performing a detailed characterization of the spatio-temporal profile of two-point density correlations in the thermodynamic limit. We actually show the compatibility of both sets of observations, unveil how the persisting ballistic expansion of the correlation front is unambiguously influenced by the chaotic phase, and we provide a deeper and more insightful understanding of correlation transport in this many-body system. 

The remainder of the manuscript is structured as follows. In Section \ref{sec:Model-Num-Meth}, we introduce the physical model under study, define the figures of merit used to characterize its dynamical behaviour and describe the numerical method employed. In Section \ref{sec:CTD}, we describe the behaviour of the two-particle correlation transport distance in the thermodynamic limit. The scrutiny of the correlation profiles in space and time is presented in Sections \ref{sec:Pseudo-Distr} and \ref{sec:Corr-Profs}, while Sections \ref{sec:velocity} and \ref{sec:decaymax} discuss chaos-induced changes in the velocity of the correlation front and its decay with distance. Finally, we summarize our findings and conclude in Section~\ref{sec:Concl}.

\section{Model and numerical method}\label{sec:Model-Num-Meth}
    \subsection{Hamiltonian and dynamical observables}
    We model a system of $N$ bosons in an $L$-site one-dimensional optical lattice using the standard Bose-Hubbard Hamiltonian (BHH) \cite{Fisher1989,Lewenstein2007,Bloch2008,Cazalilla2011,Krutitsky2016}
    \begin{align}
       H &= -J\sum^{L-1}_{j = 1} \pqty{b^\dagger_j b_{j+1} + b^\dagger_{j+1}b_j} + \frac{U}{2}\sum^L_{j = 1} \hat{n}_j(\hat{n}_j - 1) \label{eq:BHH} \\
       &\equiv H_{\rm tun} + H_{\rm int}, \notag
    \end{align}
    where $b^\dagger_j, b_j$ are the bosonic creation and annihilation operators on lattice site $j$ (i.e., in the single particle Wannier basis), and $\hat{n}_j = b^\dagger_jb_j$ are the corresponding number operators. 
    $H_{\rm tun}$ describes particle tunneling between neighbouring sites with tunneling strength $J$, while $H_{\rm int}$ accounts for the repulsive on-site interactions with strength $U > 0$.
    $H_{\rm tun}$ and $H_{\rm int}$ constitute the limits of $H$ for vanishing tunneling $(J = 0)$ or interaction $(U = 0)$, respectively, and are integrable and analytically solvable if the appropriate Fock basis is used. In contradistinction, 
    when $J, U \neq 0$, the interplay between tunneling and interaction renders the BHH non-integrable, and an ergodic (chaotic) phase emerges in its excitation spectrum that  
    can be identified 
    from spectral and eigenvector features
    \cite{Kolovsky2004,Biroli2010,Kollath2010,Beugeling2014,Beugeling2015,Beugeling2015c,Dubertrand2016,Beugeling2018,DelaCruz2020,Russomanno2020,Pausch2020,Pausch2021,Pausch2022},
    as well as from the dynamics of non-equilibrium initial configurations 
    \cite{Cheneau2012,Trotzky2012,Ronzheimer2013,Meinert2014,Kaufman2016,Pausch2025,Duenas2025,Lauchli2008,Russomanno2020,Kollath2007,Takasu2020,Markovic2025}.

    We study the dynamics of the homogeneous initial Fock state with one boson per site 
    \begin{equation}
     \ket{\psi_0} = \ket{1,1,\dots,1}
    \end{equation}     
     in the thermodynamic limit, i.e., when $L = \infty$ at unit density, for varying relative tunneling strength 
    \begin{equation}
     \gamma \equiv \frac{J}{U} >0.
    \end{equation}
     The homogeneous spatial particle density of the chosen initial state entails 
     that the net mass transport across the system remains negligible in time, and the underlying dynamical complexity is reflected in  
     a $\gamma$-dependent development of correlations of purely many-particle origin. These we characterize through the experimentally accessible two-point density correlators 
     \cite{Lukin2019,Rispoli2019,Leonard2023}
    \begin{equation}
        C_{i,j}(\tau) \equiv \expval{\hat{n}_j\pqty{\tau}\hat{n}_k\pqty{\tau}}_{\psi_0} - \expval{\hat{n}_j\pqty{\tau}}_{\psi_0}\expval{\hat{n}_k\pqty{\tau}}_{\psi_0},
    \end{equation}
    between sites $i,j \in \bqty{1, L}$, and with time measured in units of tunneling time, 
    \begin{equation}
     \tau\equiv \frac{Jt}{\hbar}.
    \end{equation}
    To quantify the correlation spatial spread in time, we introduce the two-particle correlation transport distance (CTD)
    \begin{equation}
        \ell\pqty{\tau} \equiv \sum_{d = 1}^{L-1} d \expval{\abs{C_{k,k+d}\pqty{\tau}}}_k,
        \label{eq:CTD}
    \end{equation}
    where angular brackets indicate an average over all site pairs $\pqty{k, k+d}$ for a given distance $d$. This magnitude was introduced and measured experimentally in Ref.~\cite{Rispoli2019} for the disordered BHH, and studied theoretically in Refs.~\cite{Duenas2025,Markovic2025} for the regular system. The CTD can be understood as a mean value of the distance over which two-point density correlations have propagated after a given time. 
    
    In Ref.~\cite{Duenas2025}, we demonstrated analytically (and confirmed numerically from the analysis of finite systems) that the CTD grows asymptotically ballistically in time in the two integrable limits of the BHH.   
    The onset of the chaotic phase, identified at $\gamma\approx 0.11$, slows down correlation transport 
    leading to a sub-ballistic increase of $\ell(\tau)$ compatible with a diffusive process. 
    For finite systems, the CTD eventually saturates to an $L$- and $\gamma$-dependent value. In this saturation regime, we observed that the time averaged CTD develops a non-analyticity in the thermodynamic limit when entering the chaotic regime, which is distinctly characterized by an 
    exponential suppression of the CTD temporal fluctuations with increasing $L$, extending over the range $0.11\lesssim \gamma \lesssim 20$. 
    The latter observation is in accordance with the dynamical implication of 
    the eigenstate thermalization hypothesis \cite{Deutsch1991,Rigol2008,Srednicki1996,Srednicki1999,Nandkishore2015,Deutsch2018}, and the reported ergodic regime is in perfect agreement with  
    the chaotic phase for the homogeneous Fock state unveiled by spectral analyses in previous works \cite{Pausch2025}, 
    albeit the upper $\gamma$ bound may still be dependent upon system size \cite{Pausch2020}.

    Here, we reveal the origin of the sub-ballistic dynamics for the CTD by scrutinizing the spatio-temporal behaviour of density-density correlations as functions of $\gamma$ for the $L=\infty$ system. We show that such a slowdown is indeed compatible with a ballistic propagation of the correlation front \cite{Cheneau2012,Lauchli2008,Barmettler2012,Despres2019}, which turns out to be also sensitive to the emergence of the chaotic phase.

    \subsection{Numerical method}
    \label{sec:NumMeth}
    Our numerical approach to studying the time evolution of the initial state $\ket{\psi_0}$ under the BHH for $L = \infty$ builds upon the infinite time-evolving block decimation (iTEBD) method  \cite{Vidal2007,Hastings2009}, a matrix product state algorithm particularly suited for the simulation of one-dimensional translationally invariant systems in the thermodynamic limit. 
    Since we are interested in the growth regime of the CTD prior to any finite size induced saturation (what we call \textit{steady growth regime}) 
    \cite{Duenas2025}, the iTEBD method provides an ideal framework for our simulations.
    
    iTEBD only requires the storage of three tensors to perform the time evolution \cite{Hastings2009}. This allows us to carry out the simulations with higher maximum bond dimension $\chi_{\rm max}$ than in our previous work for finite systems and standard time-evolving block decimation (TEBD) \cite{Duenas2025}, while consuming the same amount of computational resources. Thanks to this, we can reach longer times at no extra cost, while keeping the signal converged.

    \begin{figure}
        \centering
        \includegraphics[width=\textwidth]{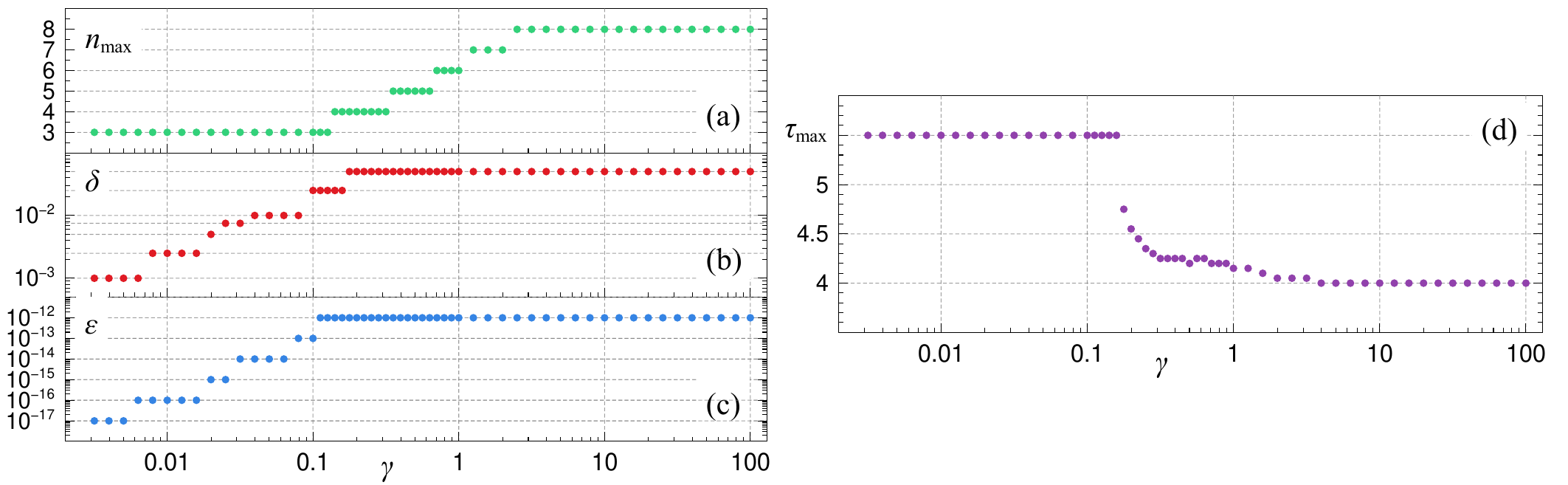}
        \caption{Optimal parameters used in the iTEBD simulations, (a) maximum local occupation number $n_{\rm max}$, (b) time step $\delta$, (c) cutoff value $\varepsilon$, and (d) maximum accessible simulation time $\tau_{\rm max}$ as functions of $\gamma$.}
        \label{fig:Sim-Params}
    \end{figure}
    
    To carry out the simulations, we use the ITensor library \cite{itensor2022,itensor-r0.32022} and implemented a custom version of the iTEBD algorithm. The local occupation number is truncated to a maximum value $n_{\rm max}$. 
    The time evolution is performed using a fourth-order Suzuki-Trotter decomposition of the evolution operator with time step $\delta$, and 
    during the evolution the bond dimension is dynamically adapted up to a maximum of $\chi_{\rm max} = 10000$ by discarding Schmidt coefficients $\lambda_n$ that satisfy
    \begin{equation}
        \frac{\sum_{n \in \mathrm{discarded}} \lambda_n^2}{\sum_n \lambda_n^2} < \varepsilon,
    \end{equation}
    where $\varepsilon$ is the cutoff value. 
    The parameters $n_{\rm max}$, $\delta$ and $\varepsilon$ are chosen for each value of $\gamma$ to ensure that the CTD is converged with a relative error below $0.5\%$ at each instant in time. The optimal values for the parameters as well as the maximum simulation times $\tau_{\rm max}$  ensuring convergence are shown as functions of $\gamma$ in Fig.~\ref{fig:Sim-Params}. Further information on our convergence criterion is given in Appendix \ref{app:numdetails}. Finally, since we are working in the thermodynamic limit, we truncate the sum in the definition of the CTD [Eq.~\eqref{eq:CTD}] to a maximum distance of $50$, which lies well beyond the front of the correlation propagation for all values of $\gamma$ and $\tau$ considered in this work.

\section{Correlation transport distance in the thermodynamic limit}
\label{sec:CTD}
    Our present results for the CTD [Eq.~\eqref{eq:CTD}] expand on those obtained in our earlier work \cite{Duenas2025} for finite systems, particularly the ones relative to 
    the steady growth regime. 
    The use of iTEBD 
    allows us to work in the thermodynamic limit ($L = \infty$ at unit density), thus eliminating finite-size effects and ensuring that the 
    time evolution remains within the steady growth regime. This, together with the ability to reach longer times, enables us to shed light over some questions that remained open in our previous study. 

    For the benefit of the reader, 
    let us first recall the CTD limiting behaviours that can be described analytically \cite{Duenas2025}. 
    For short times and $L = \infty$, an exact calculation up to order $\tau^4$ gives
    \begin{equation}
        \ell\pqty{\tau} = 4\tau^2 - \pqty{6 + \frac{1}{3\gamma^2}}\tau^4 + \order{\tau^6},
        \label{eq:Univ_Quad_Growth}
    \end{equation}
    and hence the CTD exhibits an initial universal quadratic growth. The extension of this universal regime is maximum in the non interacting limit ($\gamma \to \infty$) and shrinks as $\sim \gamma$ for strong interactions ($\gamma \to 0^+$).

    The behaviour of $\ell(\tau)$ in the thermodynamic limit can also be exactly obtained for all times near the integrable limits of the system. 
    First, in the absence of interactions, $\gamma \to \infty$, the average connected correlation admits the integral representation \cite{Duenas2025}    
    \begin{align}
        \expval{\abs{C_{k,k+d}\pqty{\tau}}}_k \underset{\gamma \to \infty}{=} \frac{2}{\pi}\int^\pi_0 J^2_0\bqty{4\tau\sin\pqty{\theta/2}} \cos\pqty{\theta d} \dd\theta, && d \geqslant 1,
        \label{eq:PseudoDistrib_gamma_inf_full}
    \end{align}
     which permits the exact computation of the CTD, 
    \begin{equation}
        \ell\pqty{\tau} \underset{\gamma \to \infty}{=} 
         4\tau^2\, \tensor[_2]{F}{_3}\pqty{\frac{1}{2}, \frac{3}{2}; 2, 2, 2; -16\tau^2},
         \label{eq:Limit_Beh_gamma_Inf_full}
    \end{equation}
    with $\tensor[_p]{F}{_q}$ the hypergeometric function. An asymptotic expansion for $\tau\to\infty$ of the latter expression yields 
    \begin{equation}
        \ell\pqty{\tau} \underset{\gamma \to \infty}{=} \frac{16}{\pi^2}\tau - \frac{\log{\tau} + 6\log{2} + \gamma_{\rm E} - 1/2}{8\pi^2\tau} + \order{\tau^{-3/2}},
        \label{eq:Limit_Beh_gamma_Inf}
    \end{equation}
    where $\gamma_{\rm E}$ is Euler's constant. 
    A similar expression is accessible 
    in the strongly interacting limit $\gamma \to 0^+$. Employing the fermionization method introduced in Ref.~\cite{Barmettler2012}, where the dynamics for strong interactions 
    are represented by propagating doublons and holons, we can write
    \begin{equation}
        \begin{aligned}
            \expval{\abs{C_{k,k+d}\pqty{\tau}}}_k &\underset{\gamma \to 0^+}{=} 4\gamma^2\Bqty{\frac{2J_1\pqty{6\tau}}{3\tau}\bqty{\frac{J_1\pqty{6\tau}}{6\tau} + \cos\pqty{\frac{\tau}{\gamma}}} + 1}, && d = 1, \\
            \expval{\abs{C_{k,k+d}\pqty{\tau}}}_k &\underset{\gamma \to 0^+}{=} \frac{4}{9}d^2\gamma^2\frac{J^2_d\pqty{6\tau}}{\tau^2}, && d \geqslant 1,
        \end{aligned}
        \label{eq:PseudoDistrib_gamma_0_full}
    \end{equation}
    with $J_d$ the Bessel function. 
    Inserting Eqs.~\eqref{eq:PseudoDistrib_gamma_0_full} in expression \eqref{eq:CTD}, leads to
    \begin{align}
        \notag 
        \ell\pqty{\tau} \underset{\gamma \to 0^+}{=} 4\gamma^2 \bigg[ 1  & +\pqty{1+48\tau^2}J^2_0(6\tau) - 8\tau J_0(6\tau)J_1(6\tau) \\
        &+ \pqty{\frac{1}{3} + 48\tau^2}J_1^2(6\tau) - \frac{J_1(6\tau)}{3\tau} 2\cos(\frac{\tau}{\gamma}) \bigg]
        \label{eq:Limit_Beh_gamma_0_full}
    \end{align}
    and brings us to the asymptotic expansion
    \begin{equation}
        \ell\pqty{\tau} \underset{\gamma \to 0^+}{=} \frac{4\gamma^2}{\pi}\bqty{16\tau + \frac{1}{6\tau} + \order{\tau^{-3/2}}}.
        \label{eq:Limit_Beh_gamma_0}
    \end{equation}
    Therefore, as we can see, the asymptotic temporal behaviour of the CTD in both integrable limits is characterized by ballistic growth, albeit with markedly different amplitudes. 

    \begin{figure}
        \centering
        \includegraphics[width=\textwidth]{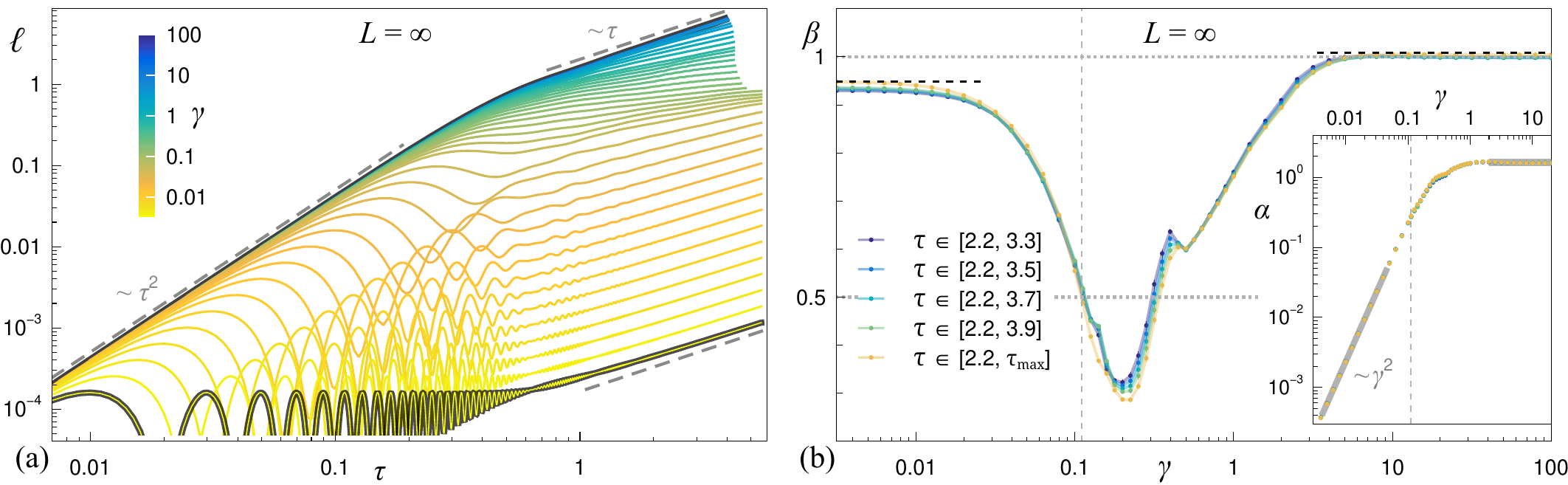}
        \caption{(a) Time evolution of the CTD for $L = \infty$ and varying $\gamma \in \bqty{0.00316, 100}$, as indicated by the color scale. The grey dashed lines highlight the initial quadratic growth [Eq.~\eqref{eq:Univ_Quad_Growth}]  and the asymptotic ballistic behaviours. 
        The values of the CTD computed analytically in the integrable limits, following Eqs.~\eqref{eq:Limit_Beh_gamma_Inf_full} and \eqref{eq:Limit_Beh_gamma_0_full}, are shown by two thicker black lines. (b) Parameters $\beta$ (main panel) and $\alpha$ (inset) of the CTD power-law fit $\ell\pqty{\tau} = \alpha \tau^\beta$, as functions of $\gamma$ for the indicated fitting intervals. Horizontal dotted grey lines highlight ballistic ($\beta = 1$) and diffusive ($\beta = 1/2$) behaviour. The vertical dashed grey line marks the onset of the chaotic phase at $\gamma = 0.11$. Horizontal dashed black lines follow from the fit of the analytical signals in the integrable limits, given by Eqs.~\eqref{eq:Limit_Beh_gamma_Inf_full} and \eqref{eq:Limit_Beh_gamma_0_full}, in the intervals $\tau \in \bqty{2.2, 3.9}$ and $\tau \in \bqty{2.2, 5.5}$, respectively. In the lower inset, all data for different $L$ collapse and the thick grey lines correspond to the coefficients of the dominant term in Eqs.~\eqref{eq:Limit_Beh_gamma_Inf} and \eqref{eq:Limit_Beh_gamma_0}. When not visible, error bars are contained within symbol size.}
        \label{fig:CTD+PowerlawFits}
    \end{figure}
    In Fig.~\ref{fig:CTD+PowerlawFits}\textcolor{blue}{(a)}, we show the behaviour of the CTD in time for $L = \infty$ and varying $\gamma$, obtained using iTEBD. The $\gamma$-dependent maximum time for which the signals are displayed, $\tau_{\rm max}\pqty{\gamma}$, is determined by our convergence criterion (refer to Section~\ref{sec:NumMeth}). As we can see, the universal quadratic growth predicted by equation \eqref{eq:Univ_Quad_Growth} is present at short times for all values of $\gamma$. As expected, the extension of this quadratic regime decreases for smaller $\gamma$. 
    In addition, the behaviour predicted by Eqs.~\eqref{eq:Limit_Beh_gamma_Inf_full} and \eqref{eq:Limit_Beh_gamma_0_full} is reproduced by the numerical signals corresponding to $\gamma = 100$ and $\gamma = 0.0031$, respectively, including the asymptotic ballistic regime anticipated by expansions \eqref{eq:Limit_Beh_gamma_Inf} and \eqref{eq:Limit_Beh_gamma_0}, already visible for $\tau>1$. As was the case with finite systems \cite{Duenas2025}, for intermediate values of $\gamma$ and the aforementioned time regime the CTD exhibits a slower, sub-ballistic growth. 

    To quantify the $\gamma$-dependence in the steady growth regime of $\ell\pqty{\tau}$, we fit the signals to a power-law of the form $\ell\pqty{\tau} = \alpha \tau^\beta$. In Fig.~\ref{fig:CTD+PowerlawFits}\textcolor{blue}{(b)}, we show the obtained fitting parameters $\alpha$ and $\beta$ as functions of $\gamma$ for $L = \infty$ and different fit (time) intervals: The initial time considered is always $\tau_{\rm i} = 2.2$ while the final time changes over the values $\tau_{\rm f} \in \Bqty{3.3, 3.5, 3.7, 3.9, \tau_{\rm max}(\gamma)}$. As observed in the inset to panel (b),
    the dependence of the coefficient $\alpha$ on $\gamma$ is not noticeably sensitive to the fit interval, and matches  
    the predictions of Eqs.~\eqref{eq:Limit_Beh_gamma_Inf} and \eqref{eq:Limit_Beh_gamma_0} in the corresponding integrable limits. 
    The exponent $\beta$ fully reveals a distinct dependence of the spreading of correlations in time on the relative tunneling strength $\gamma$: While the CTD sustains a near ballistic growth in the vicinity of the integrable limits, the onset of the chaotic phase at $\gamma\approx 0.11$ correlates with a dynamical slowdown approaching diffusion. Additionally, the $\beta$ value evolves slightly but significantly with the fit interval, as we discuss in the following.

    As $\gamma \to 0^+$, one observes $\beta<1$ in all cases. However, as the fitting interval is enlarged 
    the exponent 
    approaches $1$, consistent with the prediction of Eq.~\eqref{eq:Limit_Beh_gamma_0}. This indicates 
    that the short time scales available in the simulations are responsible for the absence of a direct observation of fully developed ballistic transport 
    in the limit of strong interactions ($\gamma \to 0^+$). One can verify this by fitting the analytical signal of Eq.~\eqref{eq:Limit_Beh_gamma_0_full} to a power-law in the time interval 
    $\tau \in \bqty{2.2, 5.5}$, 
    yielding the $\beta<1$ value indicated by the dashed black line in the upper left of Fig.~\ref{fig:CTD+PowerlawFits}\textcolor{blue}{(b)}, in full agreement with the results ensuing from the numerical signal for low $\gamma$.
    In fact, by progressively extending the time range of the fit of the analytical expression 
    one sees how $\beta$ slowly converges to $1$, approaching the asymptotic behaviour described by Eq.~\eqref{eq:Limit_Beh_gamma_0}. 
    In contrast, for $\gamma \to \infty$, even the smallest fit interval studied is sufficient to observe the asymptotic ballistic behaviour predicted by Eq.~\eqref{eq:Limit_Beh_gamma_Inf}. We also show the result of fitting the analytical result of Eq.~\eqref{eq:Limit_Beh_gamma_Inf_full} to a power-law in the interval $\tau \in \bqty{2.2, 3.9}$ by a dashed black line in the upper right of panel (b), finding agreement with the numerical signals for large $\gamma$. 
    
    Lastly, in the sub-ballistic regime, we see that the oscillation of the exponent around $0.5$ observed in Ref.~\cite{Duenas2025} for finite systems, appearing after entering the chaotic regime and being suddenly interrupted at $\gamma \approx 0.4$, remain also for $L=\infty$ for all fit intervals considered. 
    Nonetheless, the value of $\beta$ in the interval $\gamma\in [0.6, 4]$ registers a slight decline when longer times are taken into account, which suggests  
    that the width of the diffusive $\gamma$ regime may increase upon extending the duration of the dynamical evolution. 
    This is consistent 
    with the hypothesis proposed in Ref.~\cite{Duenas2025}, that attributed the absence of diffusive transport throughout the entire chaotic phase (up to $\gamma\approx 20$) to insufficiently long simulation times. 
    The increase in the maximum simulation time thanks to the use of 
    iTEBD, as compared to our previous study, provides evidence in this direction,  
    although even longer times would be necessary to fully clarify this issue. 

\section{The pseudo-distribution of the correlation transport distance}
\label{sec:Pseudo-Distr}
    To elucidate the origin of the different transport regimes exhibited by the CTD, we examine the associated discrete pseudo-distribution of the distance $d$, defined as 
    \begin{equation}
        G_d(\tau) \equiv \expval{\abs{C_{k,k+d}\pqty{\tau}}}_k,
        \label{eq:PseudoDistrib}
    \end{equation}
    which encodes the average magnitude of all two-point correlations at distance $d$. 
    As can be seen from Eq.~\eqref{eq:CTD}, the CTD corresponds to the first moment of the latter pseudo-distribution, and thus
    the dynamical evolution of $G_d(\tau)$ in the $d$ domain determines the temporal behaviour of the CTD. 
    The norm 
    \begin{equation}\label{eq:NormPseudoDistrib}
        \mathcal{N}\pqty{\tau} \equiv \sum^\infty_{d = 1} G_d\pqty{\tau}
    \end{equation}
    can be analytically obtained in the integrable limits of the BHH making use of 
    Eqs.~\eqref{eq:PseudoDistrib_gamma_inf_full} and \eqref{eq:PseudoDistrib_gamma_0_full}. 
    In the non-interacting limit, $\gamma \to \infty$, we compute 
    \begin{align}
        \mathcal{N} \pqty{\tau} \underset{\gamma \to \infty}{=}& 3\bqty{1 - \tensor[_2]{F}{_3}\pqty{\frac{1}{2}, \frac{1}{2}; 1, 1, 1; -16\tau^2}} \notag \\
        \underset{\phantom{\gamma \to \infty}}{=}& 3 - \frac{3}{2\pi^2\tau}\pqty{\log{\tau} + \gamma + 6\log{2}} + \order{\tau^{-3/2}}.
    \end{align}    
    Similarly, 
    for $\gamma \to 0^+$, we obtain
    \begin{align}
        \mathcal{N} \pqty{\tau} \underset{\gamma \to 0^+}{=}& 24\gamma^2\bqty{1-\cos{\pqty{\frac{\tau}{\gamma}}}\,\tensor[_0]{\tilde{F}}{_1}\pqty{2, -9\tau^2}} \notag \\
        \underset{\phantom{\gamma \to 0^+}}{=}& 24\gamma^2 + \order{\tau^{-3/2}},
    \end{align}    
    with $\tensor[_p]{\tilde{F}}{_q}$ the regularized hypergeometric function. These expressions highlight two key features in the integrable limits. Firstly, $\mathcal{N}(\tau) \neq 1$, 
    motivating the term \textit{pseudo-distribution} when referring to $G_d(\tau)$. And secondly,  
    the norm converges to a constant value as $\tau \to \infty$. 
    Both properties persist away from the integrable limits (as can be checked numerically \cite{Duenas2025}), being generic features 
    for all values of $\gamma$. In fact, the saturation of the norm becomes clearly visible already for $\tau\gtrsim 2$ for all $\gamma$. 

    \begin{figure}
        \centering
        \includegraphics[width=\textwidth]{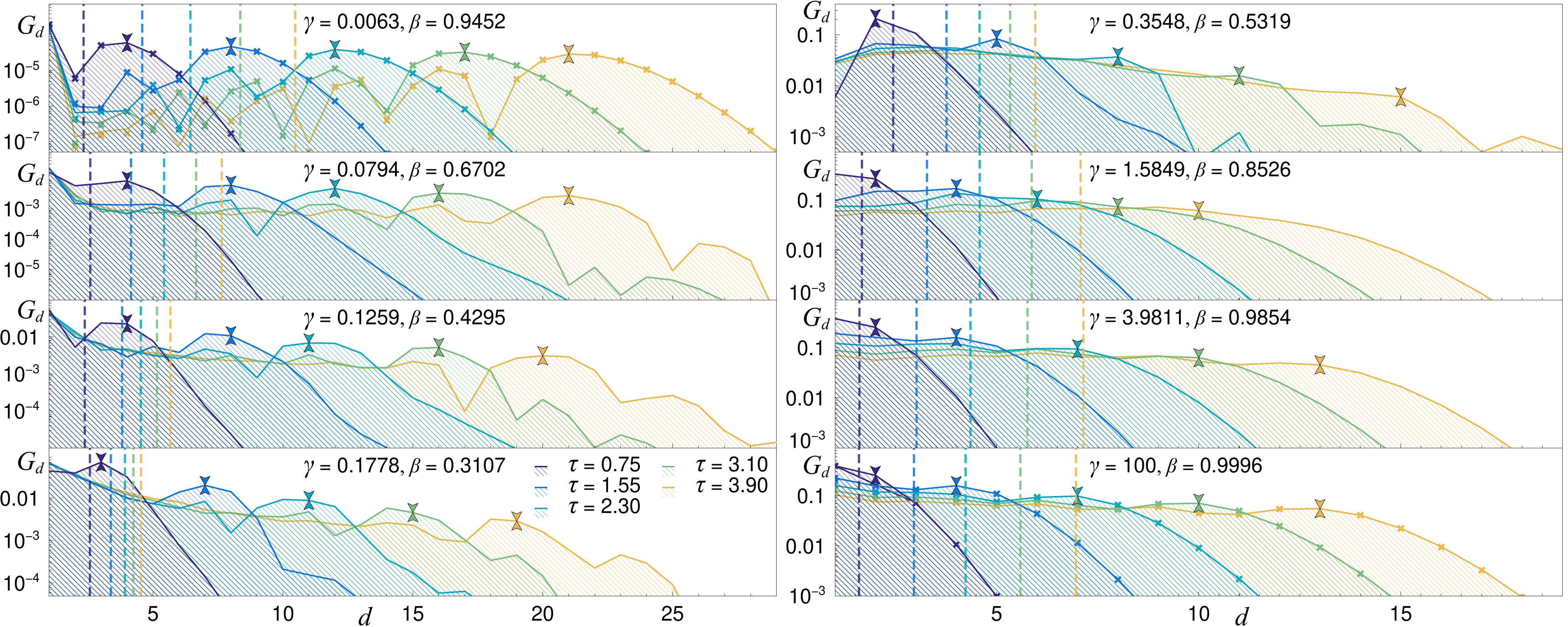}
        \caption{Pseudo-distribution $G_d\pqty{\tau}$ [Eq.~\eqref{eq:PseudoDistrib}] in log scale as a function of the distance $d$ for $L = \infty$ at five equispaced instants in time, highlighted by different colours, and different $\gamma$ values, shown in independent panels, as indicated, and accompanied by the corresponding power-law fit exponents $\beta$ [see Fig.~\ref{fig:CTD+PowerlawFits}\textcolor{blue}{(b)}]. Crosses in the panels for the lowest and highest $\gamma$ 
        display the analytical results 
        in the integrable limits, Eqs.~\eqref{eq:PseudoDistrib_gamma_inf_full} and \eqref{eq:PseudoDistrib_gamma_0_full}. Vertical dashed lines indicate the normalized CTD $\ell_{\mathcal{N}}(\tau)$ [Eq.~\eqref{eq:normalizedCTD}] for each $\tau$ and the two vertical arrowheads mark the correlation front (see main text). The corresponding visualization of $G_d\pqty{\tau}$ in linear scale can be found in Fig.~\ref{fig:Corr-Decay-Lin} in Appendix \ref{app:linlinplot}.}
        \label{fig:Corr-Decay}
    \end{figure}
    In Fig.~\ref{fig:Corr-Decay}, we show $G_d(\tau)$ as a function of the distance $d$ for $L = \infty$ at five equispaced instants in time indicated by different colours, and different $\gamma$ values shown in independent panels.  
    The figure also displays the normalized CTD, 
    \begin{equation}
    \ell_{\mathcal{N}}(\tau) \equiv \frac{\ell(\tau)}{\mathcal{N}(\tau)},
    \label{eq:normalizedCTD}
    \end{equation}
    with vertical dashed lines, and the position of the correlation front, indicated by arrowheads. 
    Note that, since 
    $\mathcal{N}(\tau)$ saturates to a constant value rather quickly, $\ell_\mathcal{N}(\tau)$ will exhibit the same power-law exponent in the steady growth regime as $\ell(\tau)$.
    The correlation front tracks the evolution of the maximum value of $G_d(\tau)$ over time for each $d$. 
    Hence, the arrowheads in the figure mark the values of $d$ for which the correlation front is closest to the time $\tau$ under consideration (the full spatio-temporal profile of the two-point correlations is shown in Fig.~\ref{fig:DenPlot+CorrProf} and will be discussed in the forthcoming section). 
    The analytical results for the integrable limits, given in Eqs.~\eqref{eq:PseudoDistrib_gamma_inf_full} and \eqref{eq:PseudoDistrib_gamma_0_full}, are also included for comparison (for $\gamma = 100$ and $\gamma = 0.0063$, respectively). 
    
    In the strongly interacting limit, 
    represented by $\gamma = 0.0063$ in Fig.~\ref{fig:Corr-Decay}, 
    the pseudo-distribution exhibits 
    a rather dominant maximum in $d$ that coincides with the correlation front. Both propagate to larger distances linearly in time, as revealed by the equispaced positions of the arrowheads for the different equispaced times considered. Since such dominant maximum carries a sizeable part of the distribution, consequently its first moment, i.e., the normalized CTD, spreads also ballistically in time (see the equally space vertical dashed lines), consistently with the value $\beta\approx 1$ identified earlier in this regime. Note that, beyond the correlation front, one should expect an exponential decay of $G_d$ with $d$ \cite{Natu2013}, which is also observed. 
    
    As $\gamma$ increases, 
    the importance of the previously dominant maximum diminishes, and the overall value of the distribution prior to such maximum rises. In fact, as time progresses, the distributions seem to coalesce into a single decaying curve with $d$ (see panels for $\gamma=0.1259$, $0.1778$, $0.3548$). Hence, although the position of the correlation front keeps advancing linearly in time, it does not carry a significant weight of the distribution, whose bulk actually registers a distinct slowdown in its evolution, clearly visible by the fact that the vertical dashed lines are no longer equally spaced, i.e., sub-ballistic spreading of the CTD, in agreement with the exponents $\beta<1$ numerically observed. 
    
    Upon further increase of the relative tunneling strength, the pronounced decay of the distribution with $d$ ceases and $G_d$ acquires an almost steady value, which decreases with time, for all distances below the correlation front (see panels for $\gamma=1.5849$, $3.9811$, $100$). This naturally leads to an acceleration of the distribution's bulk propagation, and hence of the CTD, whose evolution recovers the  linear dependence in time (note the equally spaced dashed lines), as also confirmed by the values of $\beta$ progressively approaching $1$, as expected in the non-interacting limit. As observed, in this $\gamma$ regime the correlation front keeps travelling ballistically. 
    
    Our analysis reveals two key findings. First, the correlation front propagates ballistically in time regardless of the value of $\gamma$,    
    and hence the evolution of two-point correlations exhibits a 
    propagation light-cone in $(d,\tau)$ space irrespective of the presence or not of the chaotic phase. 
    This observation agrees with results from previous theoretical \cite{Lauchli2008,Barmettler2012,Despres2019,Natu2013} and experimental studies \cite{Cheneau2012}. Secondly, despite the apparently featureless behaviour of the correlation front, the overall time evolution of the pseudo-distribution $G_d(\tau)$ is unambiguously sensitive to the emergence of quantum chaos. Upon entering the chaotic phase, the propagation of the distribution's bulk undergoes a slowdown observable in the temporal behaviour of its mean value, i.e., the CTD, which becomes sub-ballistic and nearly diffusive [cf. Fig.~\ref{fig:CTD+PowerlawFits}\textcolor{blue}{(b)}]. Interestingly, this latter deceleration is compatible with the ballistic spreading of the correlation front, although one may anticipate that, to ensure the consistency of the distribution's norm (recall that it saturates quickly in time), the magnitude of the correlation front should undergo an enhanced suppression in time within the chaotic phase. In the following sections, we unveil the imprint of the chaotic phase on the correlation front.

\section{Spatio-temporal correlation profiles}
\label{sec:Corr-Profs}
\begin{figure}
        \centering
        \includegraphics[width=.7\textwidth]{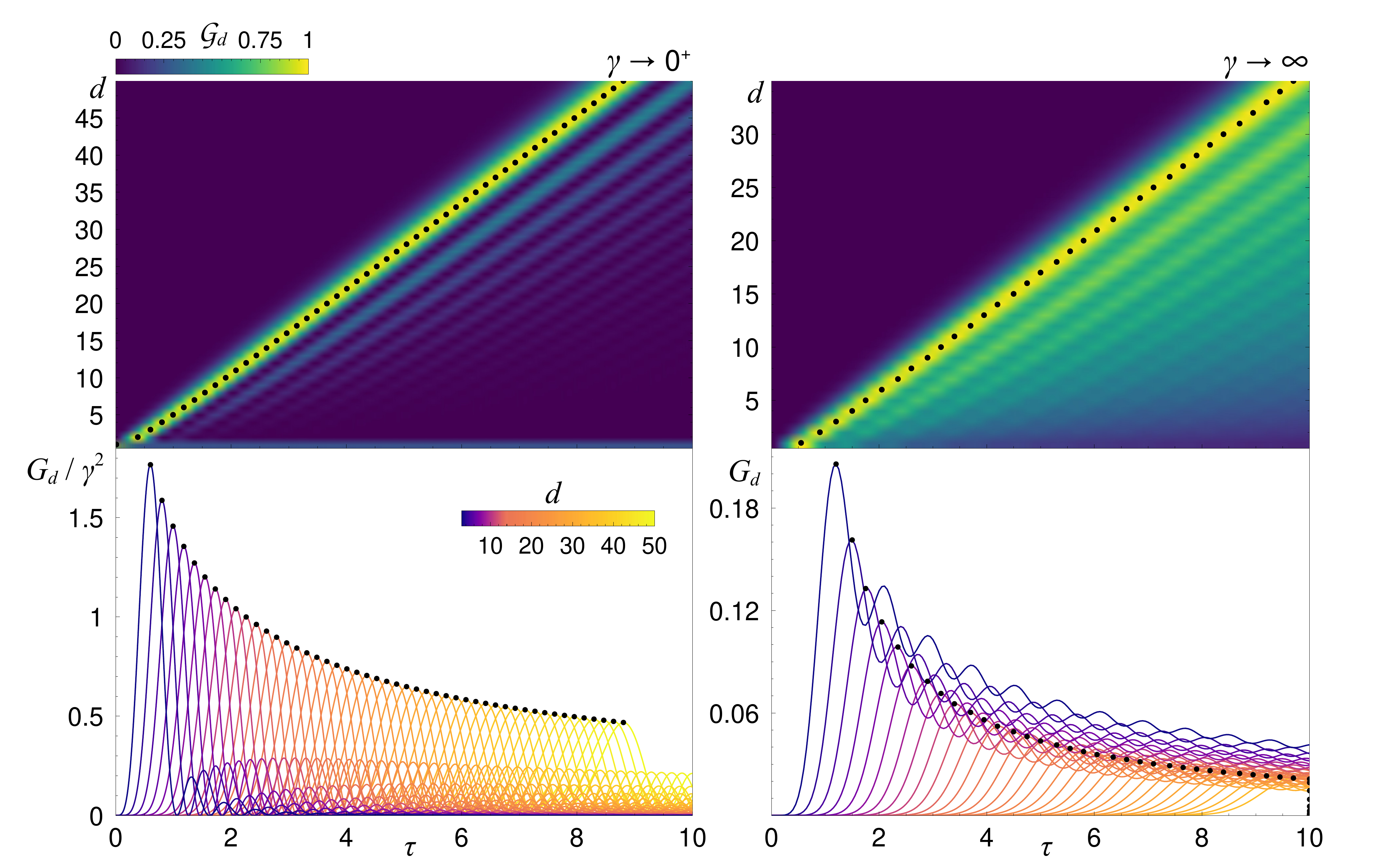}
        \caption{Analytical spatio-temporal correlation profiles for $L=\infty$ in the integrable limits $\gamma\to\infty$ (right) and $\gamma\to0^+$ (left), following Eqs.~\eqref{eq:PseudoDistrib_gamma_inf_full} and \eqref{eq:PseudoDistrib_gamma_0_full}, respectively. The upper panels display a density plot of the normalized correlation $\mathcal{G}_d(\tau)$ [Eq.~\eqref{eq:normalizedGd}] versus distance $d$ and time $\tau$. The corresponding lower panel shows $G_d(\tau)$ as a function of time for different $d\geqslant 3$ as indicated by the color scale. Black dots mark the correlation front [maximum of $G_d(\tau)$ over time for each $d$].}
        \label{fig:DenPlot+CorrProf-analytical}
    \end{figure}

\begin{figure}
        \centering
        \includegraphics[width=\textwidth]{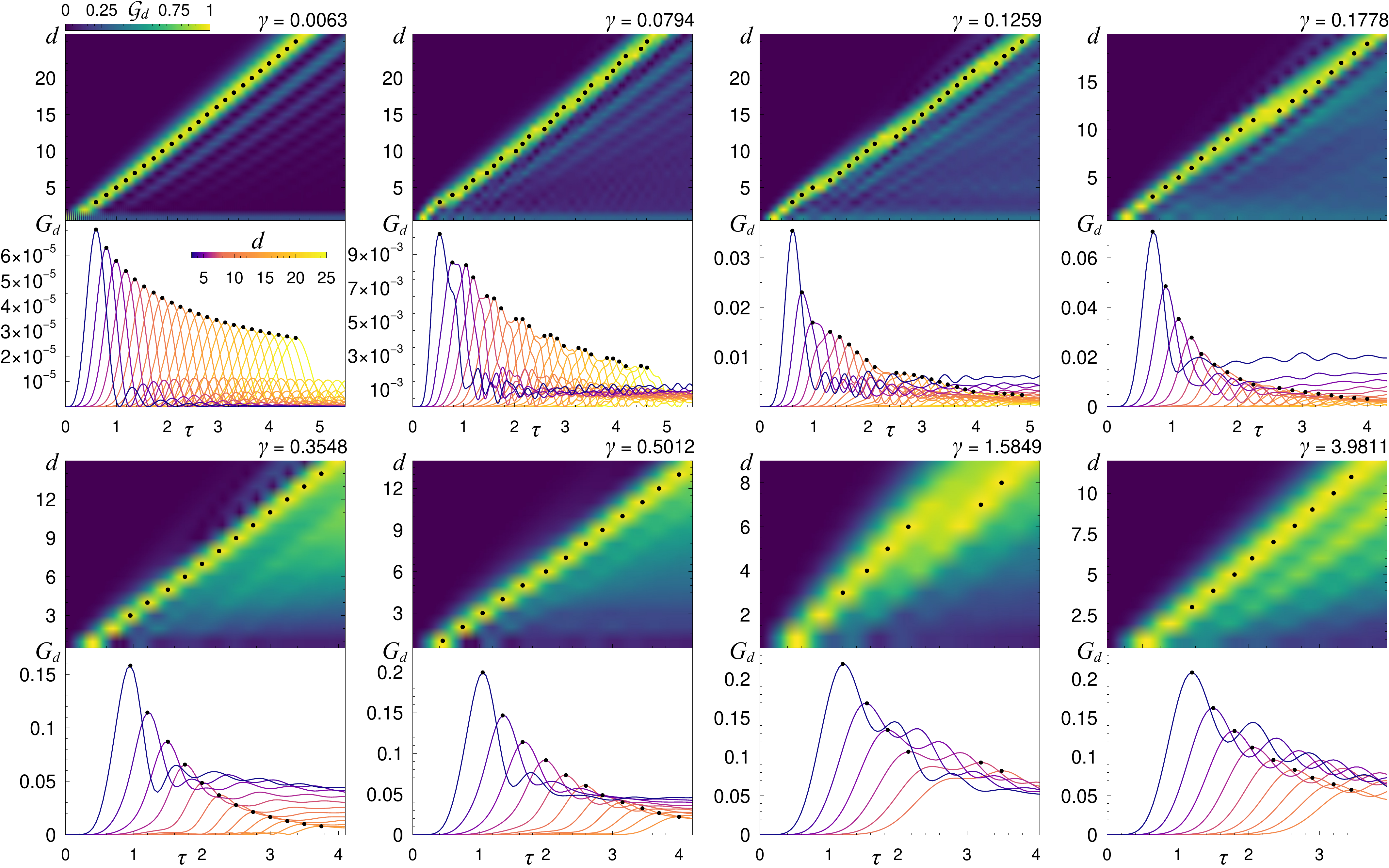}
        \caption{Numerically obtained spatio-temporal correlation profiles for $L=\infty$ and different $\gamma$ as indicated. For each $\gamma$, the upper panel displays a density plot of the normalized correlation $\mathcal{G}_d(\tau)$ [Eq.~\eqref{eq:normalizedGd}] versus distance $d$ and time $\tau$. The corresponding lower panel shows $G_d(\tau)$ as a function of time for different $d\in[3,25]$ as indicated by the color scale. Black dots mark the correlation front [maximum of $G_d(\tau)$ over time for each $d$].}
        \label{fig:DenPlot+CorrProf}
    \end{figure}

    We now turn to examining the full spatio-temporal profile of the average two-point density correlation and its dependence on $\gamma$ considering two perspectives. 
    To highlight the behaviour of the correlation light-cone, 
    we study the evolution in $(d,\tau)$-space of the normalized correlations
    \begin{equation}
        \mathcal{G}_d\pqty{\tau} \equiv \frac{G_d\pqty{\tau}}{\max_\tau\bqty{G_d\pqty{\tau}}},
        \label{eq:normalizedGd}
    \end{equation}
    where $\max_\tau\bqty{G_d\pqty{\tau}}$ denotes the maximum value of $G_d(\tau)$ over the simulated times for each $d$, i.e., the signal is normalized with respect to the value at the correlation front. 
    Additionally, 
    we look directly at the time evolution of $G_d(\tau)$ for different distances $d$. 
    These two complementary visualizations are given in the upper and lower panels, respectively, of the different plots shown in Fig.~\ref{fig:DenPlot+CorrProf-analytical}, corresponding to the analytically obtained behaviour in the integrable limits, and in Fig.~\ref{fig:DenPlot+CorrProf}, showing the numerically obtained profiles for different $\gamma$ values. 
    
    We first note that, as concluded in the previous section, the propagation of the correlation front is ballistic for any value of the relative tunneling strength, albeit certain $\gamma$-dependent novel features become also visible in the light-cone profile.  
    
    For strong interactions, represented by $\gamma=0.0063$ in Fig.\ref{fig:DenPlot+CorrProf}, the observed behaviour essentially corresponds to the one in the integrable limit $\gamma\to0^+$ shown in the left panels of Fig.~\ref{fig:DenPlot+CorrProf-analytical}: The correlation front describes a straight trajectory in $(d,\tau)$-space below which one can recognize alternating fringes of intensity, caused by the coherent oscillation in time exhibited by $G_d(\tau)$ for any $d$, expected to vanish asymptotically, according to Eq.~\eqref{eq:PseudoDistrib_gamma_0_full}. In this case, the magnitude of the correlation front is also seen to decay smoothly with $d$. 
    
    As $\gamma$ increases, and one enters the sub-ballistic regime for the CTD, i.e., approaching the onset of the chaotic phase, we observe that periodic dislocations in the light-cone structure emerge, visible in the plots for $\gamma = 0.0794$, $0.1259$, $0.1778$, although the overall linear trend is maintained. From the time evolution of $G_d(\tau)$, it is evident that such dislocations reflect a perturbation of the entire correlation profile for all $d$. The dislocations in the correlation front are first visible in the range $\gamma\in[0.016,0.2]$, where their period grows with $\gamma$, until presumably it exceeds the accessible time window. Simultaneously, in the sub-ballistic regime for the CTD, we observe how the interference pattern below the light-cone progressively loses contrast (see density plots for $\gamma = 0.0794$, $0.1259$, $0.1778$, $0.3548$, $0.5012$). In fact, here the correlations $G_d(\tau)$ do not decay to zero in time, but develop a rather prominent steady tail with a non-zero value that is inversely proportional to $d$ (e.g., see lower panels for $\gamma = 0.1259$, $0.1778$, $0.3548$). These non-zero saturation values in $G_d(\tau)$ are directly responsible for the emergence of the nearly time-independent form of the pseudo-distribution as a function of $d$ discussed in the previous section, and hence for the slowdown of the CTD evolution.

    For larger $\gamma$, we observe that the dislocations in the correlation front reappear in the range $\gamma \in [1.259,3.162]$ (see panel for $\gamma = 1.5849$ in Fig.~\ref{fig:DenPlot+CorrProf}), although now exhibiting an approximately constant period. Further $\gamma$ increase, towards the recovery of the CTD ballistic dynamics, takes the correlation profiles closer to the non-interacting limit ($\gamma\to\infty$, shown in the right panels of Fig.~\ref{fig:DenPlot+CorrProf-analytical}), where dislocations are absent, the interference pattern below the light-cone is progressively restored and $G_d(\tau)$ is expected to decay asymptotically to zero in time, as follows from 
    Eq.~\eqref{eq:PseudoDistrib_gamma_inf_full}.

    The emergence of non-zero steady values in time with suppressed temporal fluctuations, as observed for $G_d(\tau)$, is expected to be a characteristic feature for the expectation values of local observables in the chaotic regime \cite{Pausch2025,Duenas2025}. Let us then analyze the dependence of the correlation temporal tails on $\gamma$ and $d$. 
    \begin{figure}
        \centering
        \includegraphics[width=\textwidth]{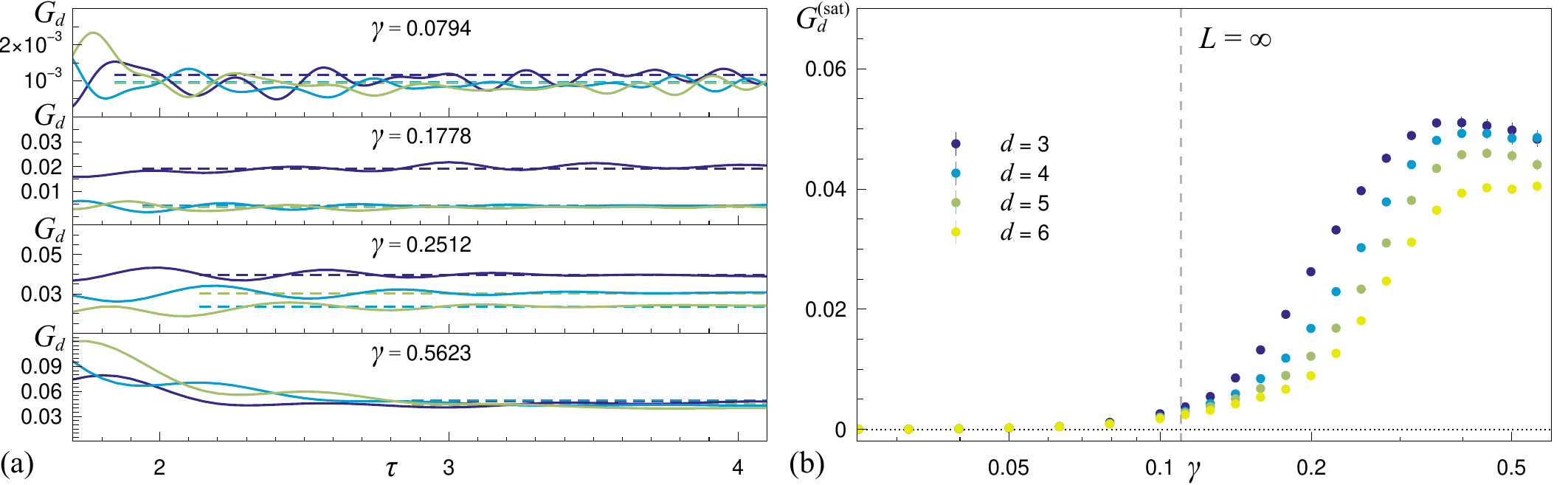}
        \caption{(a) Time evolution of $G_d(\tau)$ for $L=\infty$, different distances $d$ and values of $\gamma$ (as indicated). Dashed lines represent the saturation values $G^{\rm (sat)}_d$ as defined in the main text. (b) Saturation values $G^{\rm (sat)}_d$ as a function of $\gamma$ for different $d$ (colour coded). The vertical dashed grey line marks the onset of the chaotic phase, $\gamma = 0.11$.}
        \label{fig:Corr-Sat-Val}
    \end{figure}
    We quantify the saturation value $G^{\rm (sat)}_d$ of the correlation tail as the average of $G_d(\tau)$ over the time interval that begins one full width at half maximum after the correlation front and ends at $\tau_{\rm max}(\gamma)$. In Fig.~\ref{fig:Corr-Sat-Val}\textcolor{blue}{(a)}, we display the computed 
    $G^{\rm (sat)}_d$ (coloured dashed lines), together with the actual numerical signals of $G_d(\tau)$ (solid lines) for four exemplary values of $\gamma$. 
    Naturally, this procedure is reasonable only if a sustained value of the correlation is observed during the accessible simulation times, and we find it applicable for $d\leqslant 6$ up to $\gamma\approx 0.6$. In panel (b), we show $G^{\rm (sat)}_d$ versus $\gamma$ for different distances $d$. In the strong interaction regime
    the saturation values are negligible for all distances, and as $\gamma$ increases and crosses the onset of the chaotic phase, at $\gamma=0.11$, $G^{\rm (sat)}_d$ rises sharply up to $\gamma \approx 0.30$. Beyond this value, a weak decay is observed. Recall that in the limit $\gamma\to\infty$ correlations for any $d$ should vanish asymptotically in time.

    Therefore, the different dynamical regimes of the CTD can be directly traced back to qualitative changes in the temporal asymptotic behaviour of the two-point density correlations. When the correlations decay to zero, as it happens near the integrable limits, the distances that contribute most significantly to the CTD increase linearly in time due to the presence of the ballistic correlation front. This leads to a linear propagation of the bulk of the pseudo-distribution and, consequently, to ballistic growth of the CTD. In contrast, when correlations saturate to non-zero values, shorter distances constitute the dominant contribution to the CTD, leading a slowdown of the CTD growth. The emergence of asymptotic non-zero correlation values is directly linked to the existence of the chaotic phase. Interestingly, we observe that the transitional regions binding the ballistic and sub-ballistic CTD regimes are characterized by the appearance of dislocations in the correlation front, affecting qualitatively the profile of $G_d(\tau)$.

\section{Velocity of correlation front}
\label{sec:velocity}
    As discussed in Sections \ref{sec:Pseudo-Distr} and \ref{sec:Corr-Profs}, the correlation front propagates linearly in time for any value of the relative tunneling strength, although the profile of the correlation light-cone develops distortions in certain $\gamma$ regimes. In this section, we analyze the propagation velocity $v_{\rm cf}$ of the correlation front as a function of $\gamma$ to check whether it is affected by the presence of the system's chaotic phase.
    To this end, we fit the position of the correlation front to a relation of the form 
    \begin{equation}
     d(\tau) = v_{\rm cf}\,\tau + d_0,
     \label{eq:vcdef}
     \end{equation}
     using the nine largest distances for which the front is visible within the simulated time interval. 
    
    One can obtain an analytical approximation for $v_{\rm cf}$ in both integrable limits using Eqs.~\eqref{eq:PseudoDistrib_gamma_inf_full} and \eqref{eq:PseudoDistrib_gamma_0_full}. 
    Following the procedure outlined in Ref.~\cite{Barmettler2012}, one may estimate the time $\tau_d$ at which the first maximum of $G_d(\tau)$ occurs for a given distance $d \gg 1$,
    \begin{align}
        \tau_d \underset{\gamma \to \infty}{\approx} &\frac{1}{4}\bqty{\pqty{2d}^{1/3}\abs{z_0} + d}, \label{eq:tau_max_gamma_inf} \\
        \tau_d \underset{\gamma \to 0^+}{\approx} &\frac{1}{6}\bqty{\pqty{\frac{d}{2}}^{1/3}\abs{z_0} + d}, \label{eq:tau_max_gamma_0} 
    \end{align}
    where $z_0 \approx -1.02$ is the maximum of the Airy function. From these expressions, one can compute the instant velocity $v_d$ of the correlation front for each distance $d$,
    \begin{align}
        v_d \underset{\gamma \to \infty}{\equiv} &\frac{1}{\tau_{d+1} - \tau_d} \approx  4\pqty{1 - \frac{2^{1/3}\abs{z_0}}{3}d^{-2/3}}, \\
        v_d \underset{\gamma \to 0^+}{\equiv} &\frac{1}{\tau_{d+1} - \tau_d} \approx 6\pqty{1 - \frac{\abs{z_0}}{2^{1/3}3}d^{-2/3}}.
    \end{align}
    Here it can be seen that the instantaneous velocity of the correlation front increases with $d$ (a feature which holds for all values of $\gamma$, as can be checked numerically), approaching the asymptotic values for $d \to \infty$
    \begin{align}
        v_\infty &\underset{\gamma \to \infty}{=} 4,  \\
        v_\infty &\underset{\gamma \to 0^+}{=} 6,
    \end{align}
    and the corresponding dimensionful velocities would be obtained after multiplying by $aJ/\hbar$, $a$ being the lattice constant. From Eqs.~\eqref{eq:tau_max_gamma_inf} and \eqref{eq:tau_max_gamma_0}, one can also obtain the average correlation front velocity $v_{d,d+D}$ over distances $d$ and $d + D$,
    \begin{equation}
        v_{d,d+D} \equiv \frac{D}{\tau_{d+D} - \tau_d},
    \end{equation}
    which 
    can be identified with $v_{\rm cf}$. 
    Since our maximum available distances in the limits $\gamma\to0^+$ and $\gamma\to\infty$ are $30$ and $12$, respectively, the velocity values that we should numerically observe in the vicinity of the integrable limits are 
    \begin{align}
        v_{\rm cf} &\underset{\gamma \to \infty}{=} v_{4,12} \approx 3.59, \label{eq:vcf_gamma_inf} \\ 
        v_{\rm cf} &\underset{\gamma \to 0^+}{=} v_{22,30} \approx 5.82.  \label{eq:vcf_gamma_0} 
    \end{align}

    \begin{figure}
        \centering
        \includegraphics[width=\textwidth]{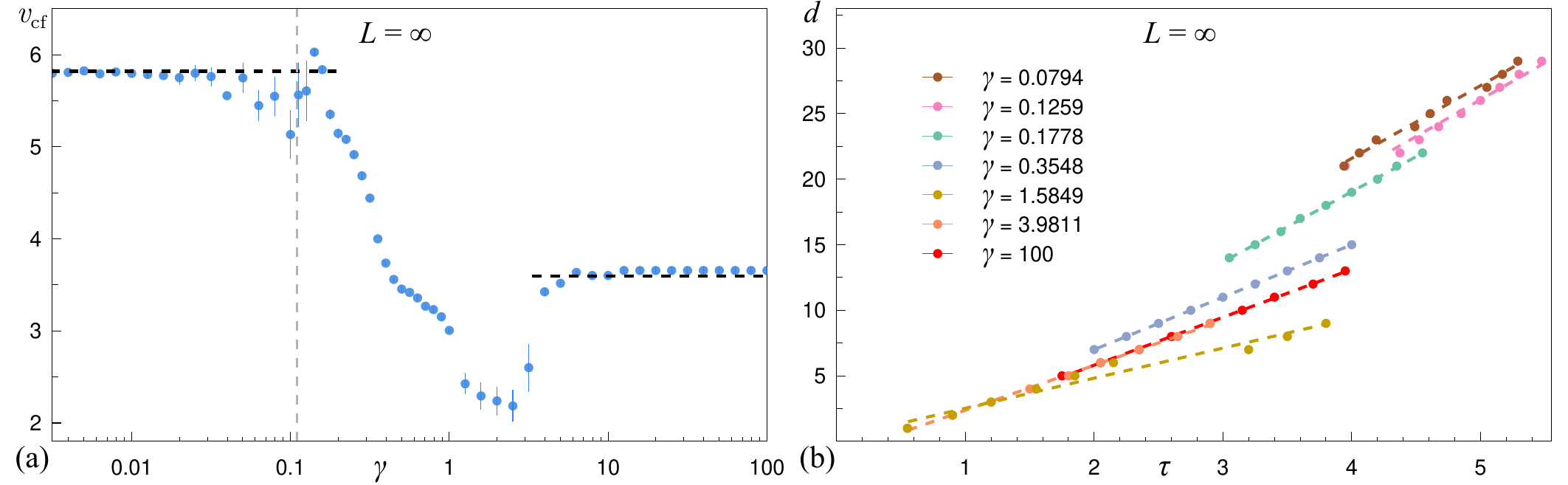}
        \caption{Velocity of correlation front for $L=\infty$. Panel (a) shows $v_{\rm cf}$ as a function of $\gamma$ ensuing from the fit to Eq.~\eqref{eq:vcdef} of the largest nine $d$ values available for each $\gamma$. Horizontal dashed black lines mark the analytical approximations in the integrable limits, Eqs.~\eqref{eq:vcf_gamma_inf} and \eqref{eq:vcf_gamma_0}. The vertical dashed grey line indicates the onset of the chaotic phase, $\gamma = 0.11$. A few exemplary cases of the linear fits are given in panel (b).}
        \label{fig:Corr-Vel}
    \end{figure}
    
    In Fig.~\ref{fig:Corr-Vel}\textcolor{blue}{(a)} we show the numerically obtained values of $v_{\rm cf}$ as a function of $\gamma$ for the simulated data, and a few linear fits using Eq.~\eqref{eq:vcdef} for some exemplary $\gamma$ values are given in panel \textcolor{blue}{(b)}. In the strongly interacting limit, the numerical estimation agrees perfectly with the analytical prediction given by Eq.~\eqref{eq:vcf_gamma_0}. When  $\gamma$ reaches the regime in which dislocations in the light-cone emerge ($\gamma\approx0.02$), the uncertainty of the $v_{\rm cf}$ values, and hence their fluctuations, increases significantly. This is to be expected, since dislocations distort the pristine linear behaviour, while maintaining the average ballistic behaviour. Once the onset of the chaotic phase at $\gamma\approx 0.11$ is crossed, the propagation velocity of the correlation front declines abruptly, reaching a suppression by almost a factor of $3$ deep within the chaotic regime. In this regime, $v_{\rm cf}$ exhibits a reduced uncertainty, which gets enhanced around $\gamma\approx 2$ due to the second occurrence of dislocations in the front, as discussed in the previous section. Further increase in $\gamma$ causes the velocity to rise again suddenly, converging quickly to the value expected in the non-interacting limit [Eq.~\eqref{eq:vcf_gamma_inf}].

    Thus, we have shown how the chaotic phase leaves an imprint in the correlation front velocity, despite the nature of the correlation light-cone being unaltered. Within the $\gamma$-regions where the CTD exhibits ballistic growth, the front velocity acquires the value of the corresponding integrable limit, whereas the slowdown of the CTD also correlates with a marked reduction of $v_{\rm cf}$. 
    Furthermore, we confirm that the emergence of dislocations in the correlation profiles signals the transition between different regimes in the correlation dynamics. 
    
\section{Decay of the correlation front with distance}
\label{sec:decaymax}
  As we mentioned in Sec.~\ref{sec:Pseudo-Distr}, the observed behaviour of the pseudo-distribution within the regime of sub-ballistic CTD dynamics implies a necessary change in the decay of the correlation front with distance, which we analyze here. Like for the velocity, we consider only the nine largest values of $d$ for which the front, with amplitude $G^{\rm (max)}_d$, is visible for each $\gamma$ in our simulations, and  characterize the decay by a power-law, 
    \begin{equation}
        G^{\rm (max)}_d \sim d^{-\eta}.
        \label{eq:Max_Decay_Fit}
    \end{equation}

\begin{figure}
        \centering
        \includegraphics[width=\textwidth]{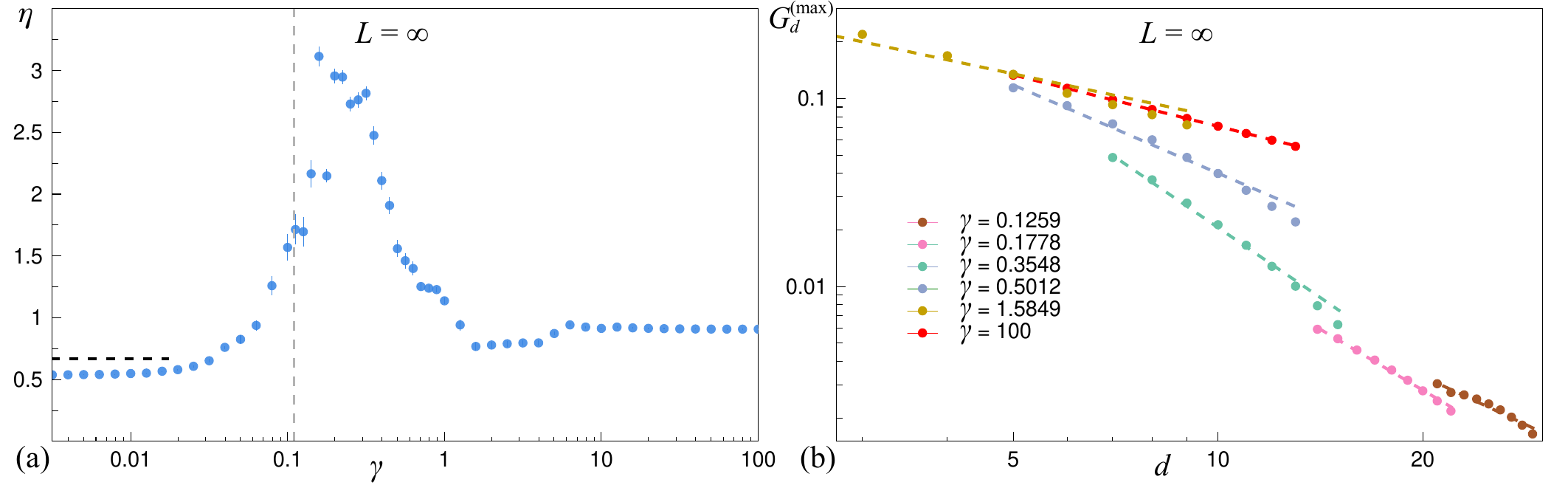}
        \caption{Decay of the correlation front with distance for $L=\infty$. Panel (a) shows the decay exponent $\eta$ as a function of $\gamma$ ensuing from the fit to Eq.~\eqref{eq:Max_Decay_Fit} of the largest nine $d$ values available for each $\gamma$. The horizontal dashed black line mark the analytical approximation in the integrable limit $\gamma\to0^+$, Eq.~\eqref{eq:Max_Decay_gamma0}. The vertical dashed grey line indicates the onset of the chaotic phase, $\gamma = 0.11$. A few exemplary cases of the power-law fits are given in panel (b).}
        \label{fig:Corr-Max}
    \end{figure}

    The latter decay function is proposed based on the behaviour observed near the integrable limits of the BHH. For $\gamma\to0^+$, it was demonstrated by means of an analytical approximation in Ref.~\cite{Barmettler2012} that the decay of the correlation front obeys 
    \begin{equation}
      G^{\rm (max)}_d \underset{\gamma\to0^+}{\sim} d^{-2/3},\quad d\gg 1.
      \label{eq:Max_Decay_gamma0}
      \end{equation}
      Similarly, the numerical investigation of the analytical form of $G_d(\tau)$ in the non-interacting limit given by Eq.~\eqref{eq:PseudoDistrib_gamma_inf_full}  reveals that 
      \begin{equation} 
	    G^{\rm (max)}_d \underset{\gamma\to\infty}{\sim} d^{-1}, \quad d\gg 1.\footnotemark
      \end{equation}
    While, in principle, there is no reason to expect the decay of the correlation front with $d$ to follow a power-law for all values of $\gamma$, we find that Eq.~\eqref{eq:Max_Decay_Fit} serves to capture correctly the tendency of the decay, as can be seen in the examples shown in Fig.~\ref{fig:Corr-Max}\textcolor{blue}{(b)}, and hence to identify potential qualitative changes when varying $\gamma$.
    \footnotetext{According to Ref.~\cite{Barmettler2012}, the decay of the correlation front in the non-interacting limit should follow the law $G^{\rm (max)}_d \underset{\gamma\to\infty}{\sim} d^{-4/3}$. However, we found that the approximations made in their derivation do not hold, and we have confirmed numerically from the analytical result for the correlation function in this limit [Eq.~\eqref{eq:PseudoDistrib_gamma_inf_full}] for $d\gg1$ that the front decays as $d^{-1}$.} 

    In panel \textcolor{blue}{(a)} of Fig.~\ref{fig:Corr-Max}\textcolor{blue}, we show the numerically found dependence of the decay exponent $\eta$ with $\gamma$. For the smallest values of $\gamma$, $\eta$ has a value close to $2/3$, in agreement with the analytical approximation discussed above. Deviations from the exact value are likely due to the finite distances used in the fits. 
    As $\gamma$ approaches and crosses the onset of the chaotic phase, 
    $\eta$ registers a sharp rise, amounting to a fivefold increase with respect to the strong interaction limit, indicating a faster decay of the correlation front with distance. The enhancement in $\eta$ persists until $\gamma\approx 2$, beyond which the exponent stabilizes around the value $1$, expected in the non-interacting limit. 
    
    We therefore observe that the emergence of the chaotic phase entails an unambiguous acceleration of the decay of the magnitude of the correlation front with distance, i.e., along the light-cone profile. 
    The $\gamma$ range where this decay is accelerated essentially coincides with the sub-ballistic regime for the CTD. In fact, the position of the maximum value of $\eta$ corresponds to the occurrence of the slowest CTD growth [cp.~Figs.~\ref{fig:Corr-Max}\textcolor{blue}{(a)} and \ref{fig:CTD+PowerlawFits}\textcolor{blue}{(b)}].

\section{Conclusions}
\label{sec:Concl}
  We have characterized the dynamical behaviour of two-point density correlations in the Bose-Hubbard model across the chaotic regime, making use of iTEBD to access the system in the thermodynamic limit at unit density. This has enabled us, through the study of the correlation transport distance (CTD) [Eq.~\eqref{eq:CTD}] and its underlying structure, to perform a substantially deeper and more insightful analysis of correlation transport that goes beyond simple light-cone pictures. Our results could be experimentally checked with cold atoms in optical lattices, following the dynamical evolution of a homogeneous density Fock state over a few tunneling times. 

  The analytical and numerical examinations for the infinite system confirms that the integrable limits of the model are characterized by a ballistic growth of the CTD, while the emergence of the chaotic phase is accompanied by a slowdown of correlation propagation manifested in a sub-ballistic behaviour, arguably compatible with diffusion, of the CTD. 
  This chaos-induced deceleration can be understood from the temporal asymptotic evolution of average correlations, $G_d(\tau)$ [Eq.~\eqref{eq:PseudoDistrib}]: Upon entrance into the chaotic regime, the correlations develop a steady non-zero value in time (with reduced temporal fluctuations) that is inversely proportional to the distance $d$. Due to these stable tails, the pseudo-distribution of $G_d$ values over $d$ acquires a very slowly varying form in time, and hence the CTD, which can be seen as the mean value of this latter distribution, registers a significant dynamical lag.
  
  Despite these observations, the two-particle correlation front is found to propagate ballistically in time for all values of the relative tunneling strength $\gamma$, i.e., irrespective of the presence or not of the chaotic phase. Nonetheless, we demonstrate that certain features of the correlation front also carry a distinctive $\gamma$ dependence: The front velocity registers a marked drop at the onset of the chaotic phase that also correlates with a faster decay of the front magnitude with distance. Additionally, the light-cone profile exhibits characteristic dislocations that seem to signal the transitional regimes between ballistic and sub-ballistic dynamics of the CTD. We thus reconcile the apparent incompatibility between the observation of diffusive correlation transport of Ref.~\cite{Duenas2025}, and earlier works 
  reporting ballistic correlation dynamics in this system \cite{Cheneau2012,Lauchli2008,Barmettler2012,Despres2019,Natu2013}.
    
  Ultimately, this work underscores the necessity of looking `beyond the light-cone' to understand the rich correlation transport phenomena in chaotic quantum systems. By establishing a direct link between the structure of two-point density correlations and the evolution of the CTD, we provide a clearer picture of how interaction-induced chaos modifies the spreading of information. These insights not only clarify previous observations, 
  but also establish the CTD and its underlying structure as powerful diagnostics for characterizing dynamical regimes, setting the stage for future investigations into the long-time persistence of 
  the observations here reported.

\begin{acknowledgements}
The authors acknowledge support through grant no. PID2024-156340NB-I00 funded by Ministerio de Ciencia, Innovaci\'on y Universidades/Agencia Estatal de Investigaci\'on (MICIU/AEI/10.13039/501100011033) and by the European Regional Development Fund (ERDF). O.D.~acknowledges support from a Ph.D Fellowship funded by European Social Fund Plus, Programa Operativo de Castilla y León and Junta de Castilla y León through Consejería de Educación. This research has made use of the high performance computing resources of the Castilla y Le\'on Supercomputing Center (SCAYLE, www.scayle.es), financed by the European Regional Development Fund (ERDF). 
\end{acknowledgements}

\newpage

\appendix
\section{Numerical method: Details on convergence criterion}
\label{app:numdetails}

To guarantee the numerical reliability of our results, 
a careful convergence analysis, resulting in the information shown in Fig.~\ref{fig:Sim-Params}, 
has been carried out with respect to the parameters needed to perform the simulations. In this Appendix, we detail such analysis.

In our previous work regarding finite systems \cite{Duenas2025}, we established the optimal parameters $n_{\rm max}, \delta$ and $\varepsilon$ needed to ensure convergence and numerical stability of the CTD signals computed using TEBD for maximum bond dimension $\chi_{\rm max} = 2500$, systems up to $L = 100$ and tunneling times $\tau \leqslant 3.3$. 
Since such temporal range is too short for the system to display any finite size effects given the lengths studied, we assumed these values to be also optimal for simulations in the thermodynamic limit employing iTEBD.
For those parameters, 
simulations using iTEBD were performed for $\chi_{\rm max} = 2500, 5000$ and $10000$ up to $\tau = 5.5$. We established the maximum time for which the CTD is converged $\tau_{\rm max}$ at $\chi_{\rm max} = 2500$ and $5000$ by comparing the corresponding signals with that obtained for $\chi_{\rm max} = 10000$. When the inequality
\begin{equation}
    \frac{\abs{\ell^{[\chi^{(i)}_{\rm max}]}(\tau)-\ell^{[\chi^{(3)}_{\rm max}]}(\tau)}}{\ell^{[\chi^{(3)}_{\rm max}]}(\tau)} \leqslant 0.005,
    \label{eq:conv_crit}
\end{equation}
no longer holds (with $i = 1,2,3$ corresponding to $\chi^{(i)}_{\rm max} = 2500, 5000, 10000$), the aforementioned convergence is considered to be broken.

\begin{figure}
    \centering
    \includegraphics[width=\textwidth]{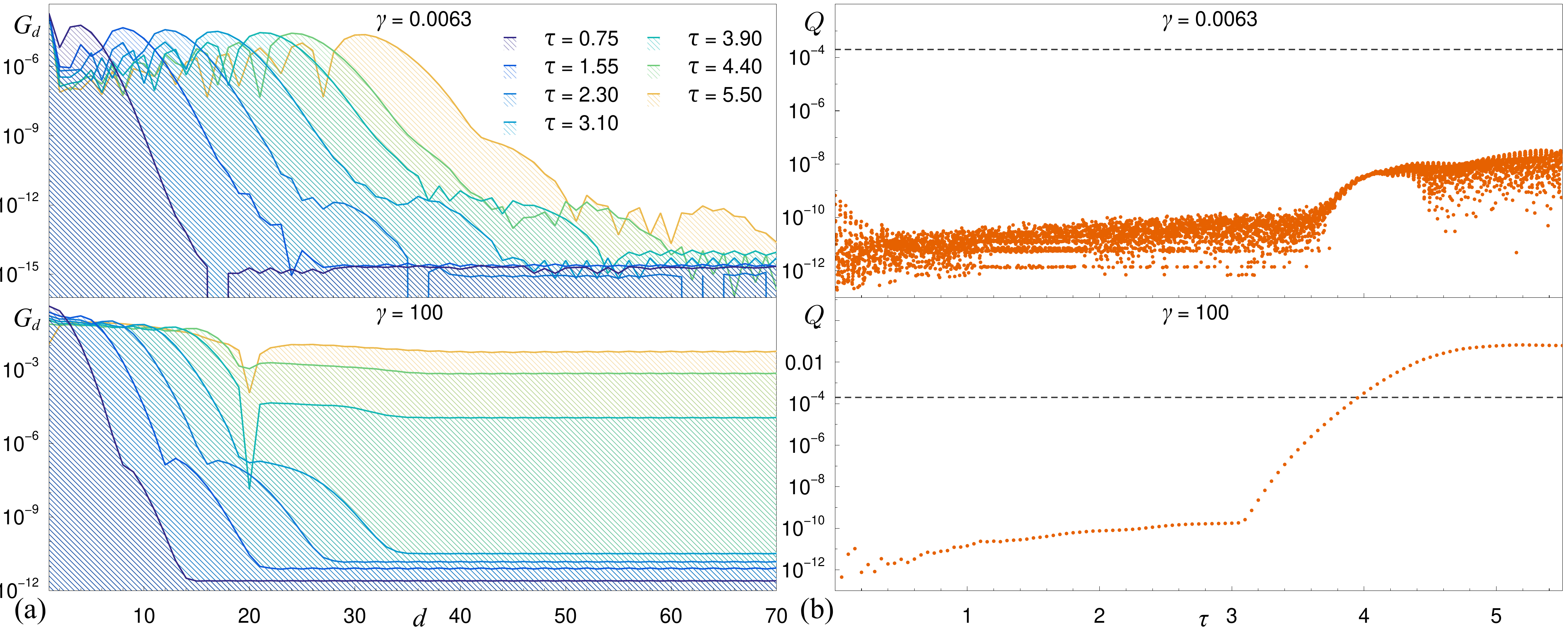}
    \caption{(a) Pseudo-distribution $G_d\pqty{\tau}$ in log scale as a function of the distance $d$ for $L = \infty$ and $\chi_{\rm max} = 10000$ at seven equispaced instants in time, highlighted by different colours. Upper panel shows $\gamma = 0.0063$, for which convergence is ensured up to $\tau_{\rm max} = 5.5$, and lower panel displays $\gamma = 100$, for which convergence is broken for  $\tau > 4.0$ [see Fig.~\ref{fig:Sim-Params}\textcolor{blue}{(d)}]. (b) $Q (\tau)$ in log scale as a function of time $\tau$ for the two values of $\gamma$ shown in (a). Horizontal black dashed lines mark the threshold value $2\times 10^{-4}$.}
    \label{fig:Corr-Decay-Appendix}
\end{figure}

To generalize this convergence criterion to the cases where the signal with twice the value of $\chi_{\rm max}$ is not accessible, we look at the dependence of $G_d(\tau)$ on the distance $d$. As can be seen in 
Fig.~\ref{fig:Corr-Decay-Appendix}, 
due to the finite precision of the numerical computations 
there is always a threshold $d$ at which the decay of the correlation stops, and beyond which the value of $G_d(\tau)$ is purely determined by numerical noise. Note that such decay after the correlation front should ideally decrease indefinitely. 
This noise base value $\sigma_G(\tau)$ informs us of the error induced in 
the correlations by the iTEBD method, and hence it seems natural to use it to generalize the convergence criterion. To do so, we monitored the time evolution of the ratio 
\begin{equation}
    Q (\tau) \equiv \frac{\sigma_G(\tau)}{\max_d{\bqty{G_d(\tau)}}},
\end{equation}
where 
$\max_d\bqty{G_d\pqty{\tau}}$ denotes the maximum value of $G_d(\tau)$ over all distances for each $\tau$, and the noise value is evaluated at a sufficiently large distance after the correlation front (in our case we estimated $d=50$ to be appropriate). Convergence is assumed to be broken at the first instant 
for which $Q(\tau) > 2 \times 10^{-4}$, which defines $\tau_{\rm max}$. One can 
check that this criterion gives values of $\tau_{\rm max}$ compatible with those obtained using \eqref{eq:conv_crit} for $\chi_{\rm max} = 2500, 5000$. Applying it to the signals with $\chi_{\rm max} = 10000$ used in the main text yields 
the values of $\tau_{\rm max}(\gamma)$ shown in Fig.~\ref{fig:Sim-Params}\textcolor{blue}{(d)}.

Once $\tau_{\rm max}$ is determined for $\chi_{\rm max} = 10000$, we double check convergence on the previously fixed parameters $n_{\rm max}, \delta$ and $\varepsilon$ by applying, in each case, a criterion analogous to that shown in \eqref{eq:conv_crit} (see Ref.~\cite{Duenas2025} for details relative to each parameter). This results in the information given 
in Fig.~\ref{fig:Sim-Params}\textcolor{blue}{(a)}, \textcolor{blue}{(b)} and \textcolor{blue}{(c)}.
 
\section{Pseudo-distribution $\boldsymbol{G_d(\tau)}$: Linear scale visualization}
\label{app:linlinplot}
\begin{figure}
        \centering
        \includegraphics[width=\textwidth]{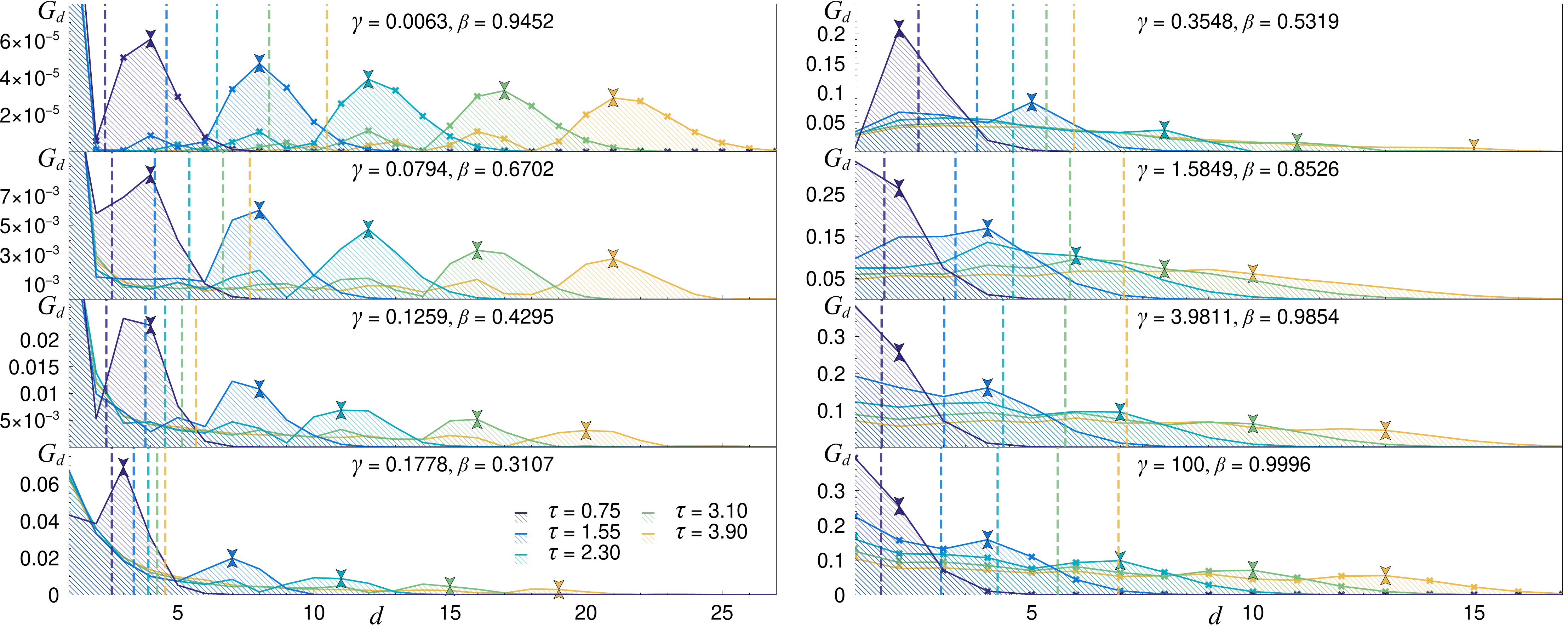}
        \caption{Representation of the pseudo-distribution $G_d\pqty{\tau}$ [Eq.~\eqref{eq:PseudoDistrib}] in linear scale as a function of the distance $d$ for $L = \infty$ at five equispaced instants in time. Colours, lines, and symbols are the same as in Fig.~\ref{fig:Corr-Decay}.}
        \label{fig:Corr-Decay-Lin}
    \end{figure}

Here, we provide an alternative visualization of the evolution of the pseudo-distribution $G_d\pqty{\tau}$ [Eq.~\eqref{eq:PseudoDistrib}] presented in Fig.~\ref{fig:Corr-Decay}. In order to highlight, in particular, the emergence, location and evolution of the correlation maxima, Fig.~\ref{fig:Corr-Decay-Lin} shows the linear scale representation of Fig.~\ref{fig:Corr-Decay}.

\providecommand{\newblock}{}

\end{document}